\documentclass[prb,twocolumn,aps]{revtex4}
\usepackage{epsfig}

\usepackage[T1]{fontenc}
\usepackage[latin1]{inputenc}
\usepackage{amsfonts}
\usepackage{amsmath}
\usepackage{epsfig}

\newcommand{\ep}{\epsilon}
\newcommand{\sig}{\sigma}

\newcommand{\Om}{\Omega}
\newcommand{\tta}{\theta}
\newcommand{\valeurabsolue}[1]{\left| #1 \right|}

\newcommand{\BDG}{Bogoliubov-de Gennes}

\begin{document}

\title{Exchange induced ordinary reflection in a single-channel
SFS junction}

\author{J\'{e}r\^{o}me Cayssol $^{1,2}$ and Gilles Montambaux $^{1}$}

\affiliation{(1)Laboratoire de Physique des Solides, Associ\'e au CNRS, Universit\'e
Paris Sud,  91405 Orsay, France}

\affiliation{(2)Laboratoire de Physique Théorique et Modèles Statistiques, Associ\'e au CNRS, Universit\'e
Paris Sud,  91405 Orsay, France}

\begin{abstract}
{The stationnary Josephson effect in a clean
Superconductor-Ferromagnet-Superconductor junction is
revisited for arbitrarily large spin polarizations. The quasiclassical calculation of the supercurrent
assumes that the Andreev reflection is complete for all channels.
However, De Jong and Beenakker have shown that the Andreev
reflection at a clean FS interface is incomplete, due to the
exchange interaction in the ferromagnet. Taking into account this
incomplete Andreev reflection, we investigate the quasiparticle
spectrum, the Josephson current and the $0-\pi$ transition in a
ballistic single channel SFS junction. We find that energy gaps
open in the phase dependent spectrum. Although the spectrum is
strongly modified when the exchange energy increases, the
Josephson current and the $0-\pi$ transition are only weakly
affected by the incomplete Andreev reflection, except when the
exchange energy is close to the Fermi energy. }
\end{abstract}
\maketitle

\today


\section{Introduction}

Ferromagnetism and singlet superconductivity are antagonist phenomena. Ferromagnetism favors spin alignment and concentrates the magnetic field lines whereas superconductivity expels the magnetic field and is supported by singlet pairing in the case of conventionnal superconductors. Nevertheless, as shown by Fulde, Ferrel, \cite{ff64} Larkin and Ovchinnikov \cite{lo65} (FFLO), superconductivity and ferromagnetism may coexist in a bulk sample for sufficiently small exchange splitting. In this case, Cooper pairs acquire a finite momentum proportional to the exchange splitting, leading to a non-uniform superconducting order parameter. However, this FFLO state has not been observed unambiguously in bulk samples. The situation is more favorable in ferromagnet/superconductor heterostructures. Owing to the proximity effect, superconducting correlations are present in the ferromagnet even in the absence of  pairing interaction. In particular, Superconductor-Ferromagnetic-Superconductor (SFS) junctions and Superconductor-Ferromagnetic-Insulator-Superconductor (SFIS) junctions can exhibit an equilibrium state where the
phase difference $\chi$ between the superconducting leads is $\pi$.  \cite{bulaev77} This so-called  $\pi$-state is reminiscent of the FFLO state. In recent experiments, the  $\pi$-state was discovered by Ryazanov {\it et al.}
\cite{ryazanov01} in SFS junctions and by Kontos {\it et al.} \cite{kontos01} in SFIS junctions. When the superconducting phase difference $\chi$ is non-zero, a non-dissipative current $I(\chi)$ flows through the junction. This so-called Josephson current is carried by
Cooper pairs in the superconducting leads and by quasiparticles in the ferromagnet. The conversion between these two kinds of carriers occurs at the interfaces by means of a scattering process known as Andreev reflection. \cite{andreev64}$^{,}$\cite{beenakker92} In the case of a clean normal metal-superconductor (NS) interface with identical Fermi velocities, an incoming spin-up electron is completely Andreev reflected as a spin-down hole, and a Cooper pair is created in the superconductor. In the presence of a tunnel barrier, the amplitude of the Andreev reflection is reduced: the incoming electron is partially reflected as a hole with opposite spin and partially as an electron with the same spin. 

De Jong and Beenakker  \cite{jong95} have studied the Andreev reflection in clean Ferromagnet-Superconductor (FS) junctions and have shown that the effect of ferromagnetism is twofold. Firstly, the exchange splitting energy $E_{ex}$ induces
a mismatch between spin-up and spin-down Fermi wavevectors. This produces an additional phase shift between electrons and holes in the ferromagnet. Secondly, in contrast to the clean NS case, the Andreev reflection is not complete: ordinary reflection appears {\it even in the absence of an insulating layer}. This phenomenon is due to the exchange potential step at the FS interface and it strongly modifies the transport
properties of a clean FS contact with a large number $N$ of modes per spin direction. As a result, the conductance of a ballistic point contact in a FS junction has been shown to decrease monotonously from $4 N e^2 /h$ in the nonferromagnetic $E_{ex}=0$ contact to zero in the half-metallic ferromagnet $E_{ex}=E_F$, $E_F$ being the Fermi energy. \cite{jong95} Using this suppression of the sub-gap conductance by the exchange interaction, an experimental method has been developped to mesure directly the spin polarization of a ferromagnetic sample by a transport mesurement.\cite{upadhyay98}$^,$\cite{soulen98} Whereas these transport properties have attracted a lot of 
theoretical and experimental interest, there are few theoretical works addressing the influence of the incomplete Andreev reflection on the thermodynamical properties of clean FS or SFS heterojunctions.\cite{halterman01}$^,$\cite{eschrig03} Indeed, the stationnary Josephson current of a clean multichannel SFS junction has been
calculated by Buzdin {\it et al.} \cite{buzdin82} in the framework of the Eilenberger equations \cite{eilen68} under
the assumption of complete Andreev reflection. The critical current has been found to oscillate as a function of the
phase shift $a=2 E_{ex} d/\hbar v_{F}$ between an
electron and its Andreev reflected hole, $d$ being the length of the ferromagnet
and $v_F$ the Fermi velocity. Moreover, due to the large number of channels, these oscillations are damped as a function of the exchange field. \cite{buzdin82}
The question arises whether incomplete Andreev reflection at a clean SFS junction may lead to a modification of the Josephson current as strong as the reduction of the conductance in a FS contact. Naively, one might expect the exchange induced ordinary reflection to have the same
physical effect as the potential barrier in a SFIS junction. In the well-studied case of SFIS junctions, ordinary reflection leads to a reduction of the Josephson current which evolves gradually towards the
usual Josephson form $I(\chi)=I_c \sin \chi$ as the transparency of the insulating layer vanishes. In the case of a short SFIS junction, Chtchelkatchev {\it et al.} have shown that the $0-\pi$ transition phase diagram depends on the transparency of the insulating layer. \cite{chtchel01} In summary, the Josephson effect has only been studied for weak ferromagnets. 

In the present paper, the thermodynamic properties of a clean single channel SFS junction are studied for {\it arbitrarily large spin polarizations}. In particular, we show how the excitation spectrum, the stationnary Josephson current, and the $0-\pi$ transition are affected by the exchange induced ordinary reflection at the FS interfaces. The paper is organized as follows: in Sec. \ref{spectrum}, we derive the phase dependent excitation spectrum of a clean SFS junction. \BDG\, equations are used in order to account for both Andreev and normal scattering. We show that the exchange induced ordinary reflection opens gaps at the phase differences $\chi=0$ and $\chi=\pi$. In comparison, we recall that there is no gap in the quasiclassical spectrum. \cite{kuple90}$^,$\cite{footquasi} In the case of a SFIS junction, gap opening occurs only at $\chi=\pi$. Sec. \ref{current} is devoted to the Josephson current, which depends on two independent parameters: the product $k_F d$, and the ratio of the exchange and Fermi energies $\eta=E_{ex}/E_F$, which mesures the spin polarization of the ferromagnet and tunes the balance between ordinary scattering and Andreev scattering at the FS interfaces. This is contrary to the quasiclassical theory in which the current is described by the single combinaison $a=2 E_{ex} d/\hbar v_F =\eta k_F \, d$. For small $\eta$, the main scattering mechanism is the Andreev reflection and the quasiclassical results are recovered in the limit $\eta \rightarrow 0$ and $k_F d \rightarrow \infty$ with finite $\eta \,k_{F}d$. For a fully polarized ferromagnet (a half-metallic ferromagnet), namely for $\eta=1$, Andreev reflection is completely suppressed, the spectrum
becomes phase independent and carries no current. In spite of the strong modifications of the spectrum, we find that the Josephson current remains almost unaffected by the exchange induced ordinary reflection up to values of the exchange field $E_{ex}$ close to $E_F$. The \, $0-\pi$\, transition is studied
in Sec. \ref{transitionpi} and is shown to be unaffected by the ordinary reflection in contrast to the $0-\pi$ transition in SFIS junctions.
\medskip


\section{Spectrum}\label{spectrum}

The excitation spectrum of a clean one-channel SFS junction is well known in the limit of very small exchange
splitting energies $E_{ex} \ll E_F$. This so-called quasiclassical spectrum is obtained by assuming that Andreev reflection is complete. With the help of the \BDG\, formalism we derive  an exact eigenvalue equation that takes into account both Andreev and normal reflection for arbitrary exchange energy $0<E_{ex}<E_F$. Even at relatively small
exchange energy, the corresponding spectrum differs from the quasiclassical one by the presence of gaps. We investigate analytically (for small spin polarization) and numerically how the Andreev spectrum evolves when the exchange energy $E_{ex}$ and the length of the ferromagnet are varied.

\subsection{Eigenvalue equation}
We consider the simplest model of a clean one-channel SFS junction. The itinerant ferromagnetism is described within the Stoner model by a one body potential $V_\sig (x)=-\sig E_{ex}$ which depends on the spin direction. The index $\sig =\pm 1$ denotes spin up and spin down. In the superconducting leads, $V_\sig (x)=0$. The kinetic part of the  Hamiltonian is

\begin{equation}
H_o = \frac{1}{2m}\left[\frac{\hbar}{i} \frac{d}{dx}-qA(x)\right]^{2} -E_F ,
\end{equation}
where $m$ is the effective
mass of electrons and holes. The vector
potential $A(x)$ is responsible for the phase difference $\chi$ between the leads, and $E_F =\hbar^2 k_{F}^{2} /2m$ is the Fermi energy. The Fermi velocities are identical in both superconductors and in the central metal for $E_{ex}$=0. When they are different ordinary and Andreev reflections are modified.\cite{zutic99}$^,$\cite{zutic00} In the absence of spin-flip scattering, the spin channels ($u_\uparrow , v_\downarrow$) and ($u_\downarrow , v
_\uparrow$) do not mix. The purely one-dimensional electron-like  $u_\sig (x)$ and hole-like wavefunctions $v_{-\sig} (x)$ satisfy two 
sets $\sig=\pm 1$ of independent \BDG\, equations

\begin{equation}\label{eqbogosfs}
\left(
\begin{array}{cc}
H_o + V_\sig (x)    &  \Delta(x) \\ \Delta(x)^* & -H_{o}^{*}+ V_\sig (x)      \\
\end{array}
\right) \left(
\begin{array}{l}
 u_\sig \\ v_{-\sig}
\end{array}
\right) =\ep(\chi) \left(
\begin{array}{l}
 u_\sig \\ v_{-\sig}
\end{array}
\right),
\end{equation}
where $\ep(\chi)$ is the quasiparticle energy mesured from the Fermi energy. The pair potential
is $\Delta(x)=\mid \Delta \mid e^{i \chi/2}$ in the left lead and $\Delta(x)=\mid \Delta \mid e^{-i \chi/2}$ in the right lead. In the central ferromagnetic segment, the pair potential is identically zero. Therefore, the eigenvectors of Eq. (\ref{eqbogosfs}) are strictly electron-like or hole-like with a plane wave spatial dependence because of the absence of disorder. The electron and hole wavevectors, denoted respectively by $k_{\ep,\eta}^{\sig}$ and $h_{\ep,\eta}^{-\sig}$, must satisfy

\begin{eqnarray}
  \frac{\hbar^2 [k_{\ep,\eta}^{\sig}]^2 }{2m} -E_F &=& \ep + \sig E_{ex}, \nonumber \\
  \frac{\hbar^2 [h_{\ep,\eta}^{-\sig}]^2 }{2m} -E_F &=& -\ep - \sig E_{ex}.
  \label{vondepre}
\end{eqnarray}
Introducing the degree of spin polarization $\eta=E_{ex}/E_F$, we obtain
\begin{eqnarray}
  k_{\ep,\eta}^{\sig} &=& k_F \sqrt{1+\sig \eta +\frac{\ep}{E_F}}, \nonumber \\
  h_{\ep,\eta}^{-\sig}&=& k_F \sqrt{1-\sig \eta -\frac{\ep}{E_F}}.
  \label{vonde0}
\end{eqnarray}
We consider only excitations the energies of which are smaller than the superconducting gap.
Matching the wavefunctions and their derivatives at the FS interfaces, we obtain the following eigenvalue equation
for the Andreev levels

\begin{eqnarray} \label{spehorrible}
16 k h  \cos \chi &=&
-2 (k^2-k_{F}^{2})(h^2-k_{F}^{2}) \left[\cos \Delta k  d -\cos \Sigma k  d \right] \nonumber\\
&-&(k-k_F )^2 \,(h+k_F )^2 \cos (\Sigma k  d + 2 \varphi_\ep) \nonumber\\
&-&(k+k_F )^2 \,(h-k_F )^2 \cos (\Sigma k  d - 2 \varphi_\ep)
\\
&+&(k+k_F )^2 \,(h+k_F )^2 \cos (\Delta k d - 2 \varphi_\ep) \nonumber\\
&+&(k-k_F )^2 \,(h-k_F )^2 \cos (\Delta k d + 2 \varphi_\ep),
\nonumber
\end{eqnarray}
where, for convenience, we define $k=k_{\ep,\eta}^{\sig}$, $h=h_{\ep,\eta}^{-\sig}$, $\Delta k =\Delta k_{\ep,\eta}^{\sig}=k - h$, $\Sigma k=\Sigma k_{\ep,\eta}^{\sig}=k + h$  and
$\varphi_\ep = \arccos(\ep/\Delta)$. The typical energies of the problem are the superconducting gap\, $\Delta$, the exchange energy\, $E_{ex}$, the level spacing in the ferromagnet\, min$(\hbar v_F /d,\Delta)$ and the Fermi energy\, $E_F$. In conventionnal s-wave superconductors, we have $\Delta/E_F<0.01$. The exact spectrum $\ep^\sig (\chi)$ depends on two dimensionless parameters: the ratio $\eta=E_{ex}/E_F$ and the product $k_F d$. In our model, the ratio $\eta$ 
is identical to the spin polarization at the Fermi level defined as ($I_\uparrow + I_\downarrow )/( I_\uparrow- I_\downarrow )$
where $I_\uparrow$ and $I_\downarrow$ are the spin-polarized currents associated respectively to spin up and down. \cite{soulen98}
The spin polarizations of strong ferromagnetic elements like Fe, Co and Ni are between $0.3$ and $0.5$. \cite{meservey73}$^,$\cite{upadhyay98} The recently discovered half metals, like La$_{0.7}$Sr$_{0.3}$MnO$_3$ and CrO$_2$, exhibit complete spin polarization.\cite{soulen98} In the present work, the spin polarization $\eta=E_{ex}/E_F$ is arbitrary and the ratio $\Delta/E_F << 1$. In a first step, we solve the eigenvalue equation (\ref{spehorrible}) perturbatively in the limit of small spin polarization
$\eta \ll 1$ for any length $d$. We complete our study by numerical results for arbitrary spin polarization in the case of short junctions.


\subsection{SNS spectrum and quasiclassical spectrum}\label{snssfsqc}
Obviously, for zero exchange field $\eta=0$, we recover the eigenvalue equation of a ballistic SNS junction \cite{kulik69}
\begin{equation}\label{specsns} \cos \chi =
\cos \left(\Delta k_{\ep,\eta=0}^{\sig} d - 2 \varphi_\ep \right)=\cos \left(
\frac{2 \ep d}{\hbar v_F} - 2 \varphi_\ep \right),
\end{equation}
with complete spin degeneracy between the ($u_\uparrow , v_\downarrow$) and ($u_\downarrow , v_\uparrow$) channels. For very small spin polarization $\eta=E_{ex} /E_F \ll 1$, a crude approximation of the equation (\ref{spehorrible}) is given by the formula \cite{kuple90} 
\begin{equation}\label{specsfsq} \cos \chi =\cos \left(
\frac{2 \ep d}{\hbar v_F}+a - 2 \varphi_\ep \right),
\end{equation}
with $a=2 E_{ex} d/(\hbar v_F)$. This expression was first obtained by solving the Eilenberger equations with a continuity assumption on both normal and anomalous quasiclassical Green's functions.\cite{buzdin82} It was also obtained later in the framework of
the linearized \BDG\, equations. \cite{kuple90}  Physically, these derivations of the SFS spectrum neglect ordinary reflection induced by the exchange potential $V_{\sig}(x)$. In this limit, the only effect of the exchange field is to modify the SNS spectrum Eq. (\ref{specsns}) by a shift $a=2 E_{ex} d /\hbar v_F =\eta k_F \, d$ of the superconducting phase. This shift lifts the degeneracy between the two spin channels ($u_\uparrow , v_\downarrow$) and ($u_\downarrow , v_\uparrow$).

\subsection{Small spin polarization}\label{smalleta}
\begin{figure}[ht!]
\begin{center}
\epsfxsize 9.0cm \epsffile{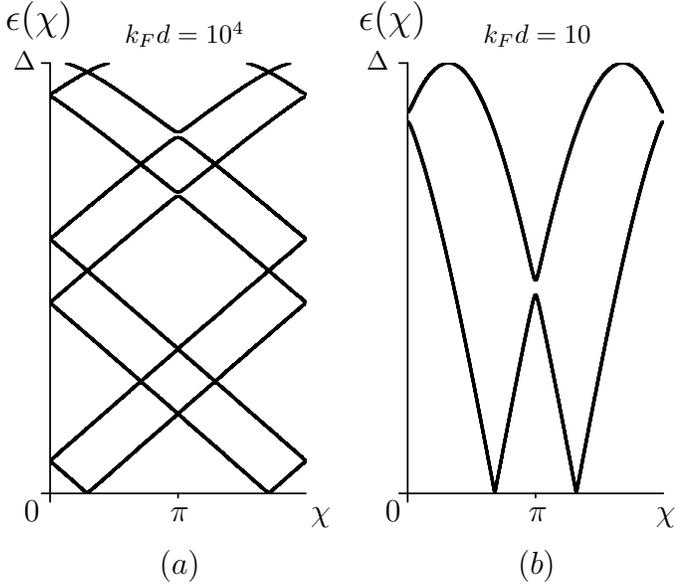}
\end{center}
\vspace*{-0.8cm}
\caption{{\it Spectrum of a clean SFS junction in the perturbative limit $\eta=0.1$ and for $\Delta/E_F=10^{-3}$. Two examples are shown: a) long junction with $k_F d = 10^4$ ($d = 5 \, \xi_o$) and  b) short junction with $k_F d =10$ ($d = 0.05 \, \xi_o$). Gaps open for $\chi=0$ and $\chi=\pi$ due to the presence of ordinary reflection. There are two zero energy Andreev levels located at the phase differences $\pi \pm \Delta k_{\ep=0,\eta} d$. }}
\label{fig1}
\end{figure}

Here, we provide a more accurate approximation of Eq. (\ref{spehorrible}). By expanding Eq. (\ref{spehorrible}) to the leading order in $\eta$, we obtain in the regime $\eta \ll 1$
\begin{equation}\label{specsfs} \cos \alpha^{\sig}(\chi,\ep) =
\cos \left(
\Delta k_{\ep,\eta}^\sig d - 2 \varphi_\ep \right),
\end{equation}
where $\alpha^{\sig}(\chi,\ep)$ is an effective phase difference related
to the true superconducting phase difference $\chi$ by the expression
\begin{eqnarray}\label{phaseeff}
\cos \alpha^{\sig}(\chi,\ep) = \left(1-\frac{\eta^2}{8}\right) \,\cos \chi &+&\frac{\eta^2}{4} \,
\frac{\ep^2}{\Delta^2} \cos 2 k_F d  \nonumber \\
&+& \frac{\eta^2}{8} \, \cos \Delta k_{\ep,\eta}^\sig d .
\end{eqnarray}
The associated spectrum depends on the length of the ferromagnet via the product $k_F d$ and
on the spin polarization $\eta=E_{ex}/E_F$. In the appendix, we calculate how this
spectrum deviates from the above mentionned quasiclassical spectrum. The largest deviations are reached for phase differences $\chi=0$\, and $\chi=\pi$\, where gaps appear, as shown in Fig. \ref{fig1}. The opening of these gaps which oscillate as a function of $k_F d$ and $\eta$ and vanish for particular values of these parameters reveals the presence of some amount of ordinary reflection. The natural energy scales for the gaps are provided by

\begin{equation}
E_\chi = \left[  \frac{d}{\hbar v_F} + \frac{1}{\sqrt{\Delta^2 - \ep_o^2 (\chi)}} \right]^{-1}.
\end{equation}
for $\chi=0$ and $\chi=\pi$ respectively.

For long junctions $d \gg \xi_o$, this energy scale is the level spacing $E_\chi \approx \hbar v_F /d$. There are many Andreev levels which cross at $\chi=0$ and $\chi=\pi$ in the non-perturbated spectrum. The amplitude of the gaps is larger in the "high-energy" spectrum close to the superconducting gap $\ep \simeq \Delta$. They vanish in the low-energy part of the spectrum $\ep \ll
\Delta$, as shown in Fig. \ref{fig1}(a). The absence of gaps at low energy is a general phenomenon, because in the limit $\ep \ll \Delta$, the eigenvalue equation Eqs. (\ref{specsfs},\ref{phaseeff}) tends to

\begin{equation}\label{speczero}
\cos \chi =\cos \left(
\frac{2 \ep d}{\hbar v_F}+\sig \frac{2 E_{ex} d}{\hbar v_F} - \pi \right),
\end{equation}
which is identical with the "gapless" quasiclassical equation Eq. (\ref{specsfsq}) because $2 \varphi_\ep \approx \pi $.

In the case of a short junction $d \ll \xi_o$, the spectrum contains only two spin-polarized Andreev levels $\sig=\pm 1$ given by
\begin{equation}\label{sfsandrealpha}
\ep_\sig (\alpha)=\Delta \valeurabsolue{\cos \left( \frac{\alpha(\chi,\ep) + \sig
a}{2} \right)} .
\end{equation}
The expressions for the gap $\delta_o$ at $\chi=0$ 
\begin{equation}\label{smallgaps1}
\delta_0 =\frac{\eta \Delta}{2} \valeurabsolue{\, \sin  k_F d \, \sin \eta k_F d},                
\end{equation}
and for the gap $\delta_\pi$ at $\chi=\pi$
\begin{equation}\label{smallgaps2}           
\delta_\pi =\frac{\eta \Delta}{2}  \valeurabsolue{\, \cos k_F d \, \sin \eta k_F d}
\end{equation}
are derived in appendix. The gaps $\delta_0$ and $\delta_\pi$ vanish simultaneously when the shift between an electron and its Andreev reflected hole is $\eta k_F d=n \pi$ with $n=...,-1,0,1,...$. When the ferromagnet length corresponds to an interger or half-integer number of Fermi wavelengths, namely when $k_F d = n \pi$, $\delta_o$ vanishes and $\delta_\pi$ is maximal. If the size of the ferromagnet and the Fermi wavelength satisfy $k_F d = (n+1/2) \pi$, one obtains the opposite configuration: $\delta_\pi$ is zero and $\delta_0$ is maximal.

It is instructive to compare these results with the case of a SFIS junctions for which the ordinary reflection originates from the potential barrier of the insulating layer.\cite{chtchel01} At the usual level of approximation, a SFIS junction is described by two parameters: the electron-hole phase
shift $a=2 E_{ex} d/\hbar v_F $ and the transparency $D$ of the insulating layer. Similarly to the case of a clean SFS junction, the spectrum is given by
\begin{equation}\label{sfsandrealpha2}
\ep_\sig (\alpha)=\Delta \valeurabsolue{\cos \left( \frac{\alpha(\chi) + \sig
a}{2} \right)} ,
\end{equation}
but the effective phase has a different expression \cite{chtchel01}

\begin{equation}\label{phaseeff4}
\cos \alpha(\chi) = 1-2D \sin^2 \frac{\chi}{2}.
\end{equation}
This effective phase leads to the gaps $\delta_0 = 0$ and $\delta_\pi = 2 \sqrt{1-D} \cos (a/2)$. There is only one gap located at $\chi=\pi$, and it is independent of $k_F d$, whereas in an exact treatment of a SFIS junction with large spin polarization $\eta$, the gaps should depend on it. In this latter case ordinary reflection would originate from both insulating layer and exchange splitting.

\begin{figure}[ht!]
\begin{center}
\epsfxsize 9.0cm \epsffile{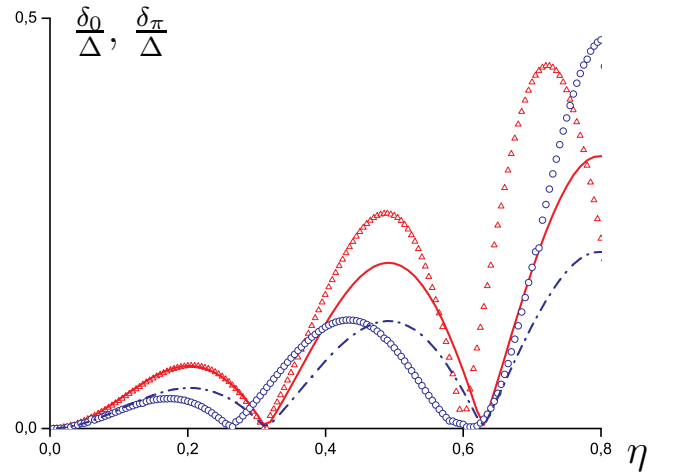}
\end{center}
\vspace*{-0.6cm}
\caption{{\it Gaps at $\chi=0$ (circles) and $\chi=\pi$ (triangles) as a function of the spin polarization $\eta$ in a short junction with $k_F d=10$. The  Eqs. (\ref{smallgaps1},\ref{smallgaps2}) provide a good approximation for small $\eta<0.1$ (dashed-dotted line: $\delta_o$, solid line: $\delta_\pi$).}}\label{figgap}
\end{figure}
In the following paragraph, we
check the validity of our results for larger exchange energies.

\subsection{Arbitrary spin polarization: numerical study}\label{arbitraryeta}

\begin{figure}[ht!]
\begin{center}
\epsfxsize 9.0cm \epsffile{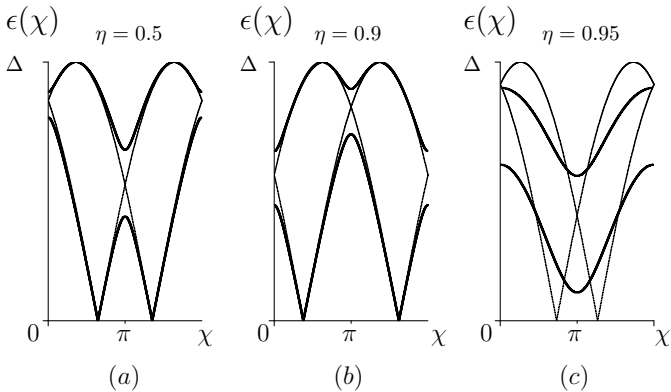}
\end{center}
\vspace*{-0.8cm}
\caption{{\it Spectrum of a short SFS junction for increasing spin polarizations $\eta=E_{ex}/E_F$
with $k_F d =10$. The thick solid lines correspond to the spectrum obtained by 
solving Eq. (\ref{spehorrible}). The thin lines represent the corresponding quasiclassical 
estimates with $a= (\sqrt{1+\eta}-\sqrt{1-\eta}) k_F d$. We have chosen $\Delta/E_F =10^{-3}$.}}
\label{fig2}
\end{figure}

For large spin polarizations $\eta=E_{ex}/E_F$, the perturbative
approach breaks down and finding the solutions of Eq.
(\ref{spehorrible}) is a harder task. In the case of a small
junction $d \ll \xi_o$, we solve Eq.
(\ref{spehorrible}) numerically and obtain the two Andreev levels $\ep^\sig
(\chi)$ for each value of the phase difference $\chi$. Typical
results are shown in Fig. \ref{fig2}\, for increasing spin 
polarizations $\eta$ and for a particular value of
$k_F d=10$. In the perturbative regime $\eta < 0.2$, it has been shown in Sec.
\ref{smalleta} that the exact spectrum is very close to the
quasiclassical spectrum except in the vicinity of $\chi=0$ and
$\chi=\pi$. Figs. \ref{fig2}(a) and \ref{fig2}(b) show that this
statement is still valid up to very large spin polarizations. But
above a particular spin polarization $\eta^*$, the spectrum
undergoes a qualitative change : the lowest Andreev level does no
longer cross the Fermi level, as shown in Fig. \ref{fig2}(c)

To understand this crossover, we calculate the superconductive
phase difference $\chi_o^\sig$ corresponding to a zero energy
Andreev state. For sufficiently small spin polarization $\eta<0.2$, it
is always defined and given by
\begin{eqnarray}
\chi_o^{\sig}&=&\pi + \Delta k_{\ep=0,\eta}^\sig d \\
&=&\pi + \sig (\sqrt{1+\eta}-\sqrt{1-\eta}) k_F d ,\nonumber
\end{eqnarray}
but close to the half-metal case $\eta \approx 1$, the eigenvalue equation (\ref{spehorrible}) leads to
\begin{eqnarray}\label{etastar}
\cos \chi_o^\sig &=& -\frac{\sin (\sqrt{1-\eta} \, k_F d) \sin (\sqrt{1+\eta} \, k_F d) }{2 \sqrt{2 (1-\eta)}}
\end{eqnarray}
which has two solutions for $\eta < \eta^* $ and no solution for $\eta > \eta^*$.

\begin{figure}[ht!]
\begin{center}
\epsfxsize 8.0cm \epsffile{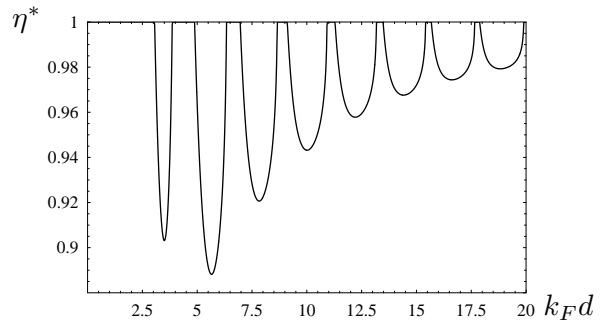}
\end{center}
\vspace*{-0.5cm} \caption{{\it  The zero-energy
Andreev states disappear above a critical polarization $\eta^*$ which 
depends on $k_F d$. In section III, we show that the current is 
very close to the quasiclassical estimate with discontinuities when the Andreev level crosses the
Fermi level [Fig. \ref{fig4}(a)] for
$\eta < \eta^*$. For $\eta > \eta^*$, the
Josephson current is strongly modified and has no discontinuity,
since a gap opens at the Fermi level [Figs \ref{fig4}(b) and 
\ref{fig5}]. The minima of $\eta^* $ correspond 
to values of $k_F d \simeq (n+1/2) \pi / \sqrt{2}$. }} \label{fig2eta}
\end{figure}

Fig. \ref{fig2eta}\, shows that the critical polarization $\eta^*$
depends on the length $d$ of the ferromagnet in a very peculiar
way. For   $k_F d < 3$, the Andreev spectrum has always two states
at the Fermi level. For\, $k_F d >3$, $\eta^*$ becomes smaller than
$1$. For spin polarizations above the critical value $\eta^*$, the Andreev spectrum has
now a gap at the Fermi level. In the next section, we will study
how this gap affects the Josephson current. Even more strikingly,
when the  length $d$ increases, the gap at the Fermi level
alternatively closes and reopens : one has  an alternance between
regions with\, $\eta^* < 1$  (such as in Fig. \ref{fig2}, a gap opens
at the Fermi level) and regions with\, $\eta^* =1$ (with no gap at
the Fermi level). Practically, for  $k_F d
> 10$, no gap opens at the Fermi level for polarizations smaller than  $\eta^*=0.94$.

\section{Josephson current}\label{current}

In this section, we  obtain the Josephson current through a clean short SFS junction for {\it arbitrary large spin polarizations}. In particular, we study how the incomplete Andreev reflection induced by the ferromagnet 
affects the current. For $\eta \ll 1$, ordinary reflection is negligible and the current is
given by the usual quasiclassical expression. In the case of a half-metal $\eta=1$, the current vanishes due to the complete suppression of the Andreev reflection. We study the crossover between these two limits by calculating the current from the spectrum obtained in the previous section.

\subsection{Josephson current}

The Josephson current is given by
\begin{equation}\label{jojo}
I(\chi)=\frac{2e}{\hbar} \frac{\partial \Om}{\partial \chi},
\end{equation}
where $\Om(\chi)$ is the phase dependent thermodynamic potential. The potential can be
calculated from the excitation spectrum by using the formula \cite{bardeen69}
\begin{eqnarray}
\Omega(T,\mu,\phi)&=& - 2T \int_{0}^{\infty} \sum_{\sig} \ln \left(2 \cosh
\frac{\ep_{\sig}(\chi)}{2T} \right) \nonumber \\ &+& \int dx
|\Delta(x)|^2 / g + Tr H_o \label{Potexcit}
\end{eqnarray}
We restrict our attention to the short junction case. For each value of $\chi$, we solve Eq. (\ref{spehorrible}) numerically to obtain the two spin-polarized Andreev levels. Then, we obtain numerically the current using Eqs. (\ref{jojo},\ref{Potexcit}).

\subsection{Quasiclassical current}

For a weak ferromagnet $\eta \ll 1$, the assumption of complete Andreev reflection is justified. Therefore, one may compute the
current from the spectrum (\ref{specsfsq}) (here for $d \ll \xi_o$) and obtain the so-called quasiclassical current \cite{buzdin82}
\begin{eqnarray}\label{currentone}
I_{qc}(\chi,a)&=&\frac{\pi \Delta}{\phi_o} \sum_{\sig=\pm 1} \sin \frac{\chi+\sig a}{2}\\
&&\tanh \left( \frac{\Delta}{2T} \cos \left(\frac{\chi+\sig a}{2}\right)\right)\nonumber .
\end{eqnarray}
Except for the presence of the phase shift 
\begin{equation}
a=(\sqrt{1+\eta}-\sqrt{1-\eta}) k_F d,
\end{equation} 
the formula (\ref{currentone}) is similar to the expression for the single mode current in a short SNS junction. \cite{KOpropre1}$^{,}$\cite{Beenakker1}
In the $T=0$ case, the current-phase relationship of a SNS junction has a sharp discontinuity at $\chi=\pi$ because the lowest Andreev level passes below the Fermi level while another Andreev level carrying an opposite current moves above. \cite{Beenakker1} In the SFS junction case, the degeneracy of the Andreev levels is lifted, and this crossing occurs respectively at $\chi^\sig=\pi+\sig a$ for each of the non-degenerate Andreev levels. Consequently, the current shows two jumps at these
phase differences, as shown in Fig. \ref{fig4}(a).

\subsection{Crossover from $\eta=0$ to $\eta=1$}

In the section \ref{arbitraryeta}, we have obtained a sharp crossover between (i) a regime where the quasiclassical spectrum is only modified by gaps opening at $\chi=0$ and $\chi=\pi$ and (ii) a regime where the Andreev spectrum is strongly modified by the vanishing of the zero energy states.

\begin{figure}[ht!]
\begin{center}
\epsfxsize 9.0cm \epsffile{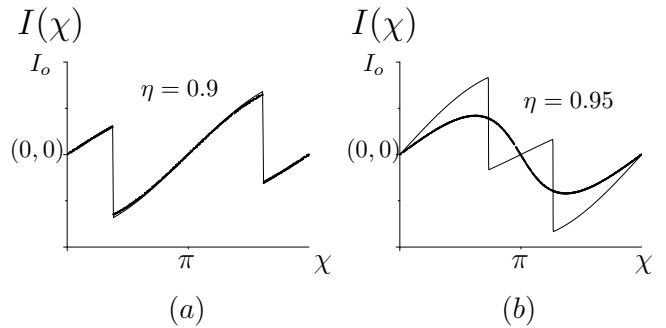}
\end{center}
\vspace*{-0.6cm}
\caption{{\it Zero-temperature current of a short SFS junction with
$k_F d =10$. a) even for a nearly complete spin polarization $\eta=0.9$, the exact current (thick line) and the quasiclassical approximation (thin line) are identical. b) for $\eta=0.95$, they are completly different. The natural scale for the current is $I_{0}=2 e \Delta/\hbar$. }}\label{fig4}
\end{figure}

For spin polarizations $0<\eta<\eta^* $, the current is well approximated
by the quasiclassical formula Eq. (\ref{currentone}) except for phase differences close to $\chi=0$ and
$\chi=\pi$. Near these values, it turns out that the correction
of the level energies induces opposite changes on the two individual currents. The sum of these corrections cancels out and the total Josephson current is unchanged. Consequently although the spectrum is modified, one may still use the quasiclassical formula (\ref{currentone}) at the current {\it for any value of $\chi$} with a very good accuracy. This statement is valid up to very high spin polarization, as shown in Fig. \ref{fig4}(a) In the limit $\eta < 0.2$, the effective phase approach leads to
\begin{equation}
I(\chi,a)=\left(1-\frac{\eta^2}{8} \right)  \, \frac{\sin
\chi}{\sin \alpha} \, I_{qc}(\alpha,a) .
\end{equation}
In conclusion, ordinary reflection induces only a very small reduction
of the current of order $\eta^2$.

When $\eta^* <\eta<1$, the current-phase relationship is
completely modified and becomes nearly sinusoidal, as shown in Fig.
\ref{fig4}(b). The discontinuity in the current disappears because
a gap opens at the Fermi level: there is no Andreev level at zero
energy.
\begin{figure}[ht!]
\begin{center}
\epsfxsize 7.0cm \epsffile{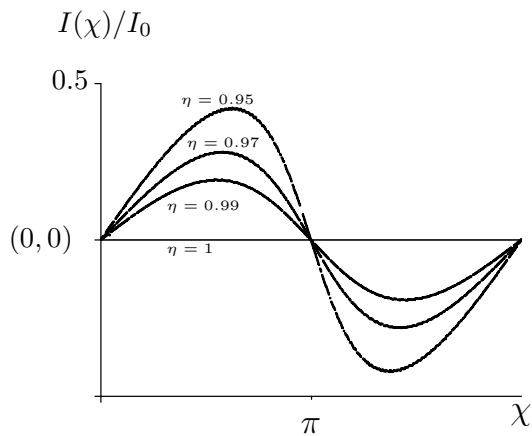}
\end{center}
\vspace*{-0.8cm}
\caption{{\it Current-phase relationships for $k_F d=10$ and various spin
polarizations in the regime $\eta^* < \eta <1$.  }}\label{fig5}
\end{figure}
In conclusion, the crossover between the regime where the current
is given by Eq. (\ref{currentone}) and the regime with zero
current $\eta > 1$, takes place in a narrow window of spin
polarizations, typically for $0.94<\eta<1$ when $k_F d=10$. For larger $k_F d$, the width of this 
window scales as $1/(k_F d)^2$.


\section{Transition $0-\pi$ in small SFS junctions} \label{transitionpi}

In this section, we study the effect of exchange induced ordinary reflection on the $0-\pi$ transition in the case of  short junctions. In order to compare the stability of the zero-phase and of the $\pi$-phase states, we compute the energy
\begin{eqnarray}\label{energ}
E(\chi,a)&=&-\Delta \sum_{\sig=\pm 1} \valeurabsolue{\cos \left( \frac{\alpha(\chi,\ep) + \sig
a}{2} \right)}
\end{eqnarray}
In the perturbative regime $\eta <0.2$, the effective phase approach applies and one obtains
\begin{eqnarray}\label{phaseeff0pi}
\cos \alpha(\chi=0) &=& (1-\frac{\eta^2}{8}) \,+ \frac{\eta^2}{8} \, \cos a  \nonumber \\
\cos \alpha(\chi=\pi)& =& - (1-\frac{\eta^2}{8}) \,+ \frac{\eta^2}{8} \, \cos a .
\end{eqnarray}
Thus
\begin{eqnarray}\label{phaseeff0pi2}
\alpha(\chi=0) &=& \pm \frac{\eta}{\sqrt{2}} \, \cos \frac{a}{2}  ,\nonumber \\
\alpha(\chi=\pi)& =& \pi \pm \frac{\eta}{\sqrt{2}} \, \sin \frac{a}{2}.
\end{eqnarray}

\begin{figure}[ht!]
\begin{center}
\epsfxsize 5.0cm \epsffile{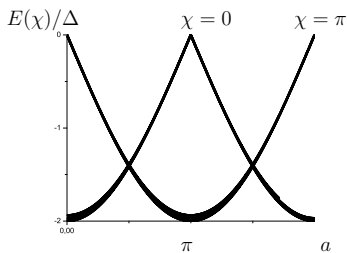}
\end{center}
\caption{{\it Zero-state energy $E(\chi=0,a)$ and
$\pi$-state energy $E(\chi=\pi,a)$ for
$\eta=0.1, 0.3, 0.5$. The intersections of the different curves remain in the vicinity of
$a=\pi/2$ and $a=3\pi/2$.}}\label{sfseffective}
\end{figure}
The energies $E(0,a)$ and $E(\pi,a)$ are represented in Fig. \ref{sfseffective}. When $E(\chi=\pi,a)>E(\chi=0,a)$, the zero-phase state is stable and the $\pi$-phase state is instable. The
curves corresponding to different values of $\eta$ are close to each other and differ slightly
only in the vicinity of $a=0$ and $a=\pi$. All these curves intersect at the same $0$-$\pi$ transition points
$a=\pi/2$ et $a=3\pi/2$. Therefore the $0-\pi$ transition is not modified by the ordinary reflection induced by the ferromagnet.

In the moderate and strong polarization regimes, numerical calculation of the energies
$E(\chi=0)$ and $E(\chi=\pi)$ as a function of\, $a=\Delta k_{\ep=0 ,\eta} = (\sqrt{1+\eta}-\sqrt{1-\eta}) k_F
d$\, leads to the same conclusion. The transition in a SFS junction at large exchange field
is robust to ordinary reflection induced by the exchange field. This is contrary to what happens in the SFIS
case. \cite{chtchel01} The energy $E(0,a)=-2 \Delta \mid \cos a \mid$ of a SFIS junction is
independent of the transparency $D$, whereas $E(\pi,a)$ evolves gradually as the transparency $D$ is varied. As a result, the transition points strongly depend on $D$: the domain of stability for the $\pi$-phase shrinks around the value
$a=\pi$ and even disappears at $D=1$. In a clean SFS junction, the stability domain
of the $\pi$-phase remains unchanged because of the interplay between the two gaps at $\chi=0$ and $\chi=\pi$. It is reminiscent
of the Josephson current robustness obtained in the previous section.

\section{Conclusion}

We have obtained the phase dependent excitation spectrum of a clean one-channel SFS junction {\it for arbitrary spin polarizations}. The present treatment takes into account the ordinary reflection of electrons caused by the ferromagnet/superconductor interface. We
have shown that gaps open for phase differences $\chi=0$ and $\chi=\pi$. These gaps depend both
on the spin polarization $\eta=E_{ex}/E_F$ and on the length of the ferromagnet via the product $k_F d$. In spite of 
these strong modifications of the spectrum, the Josephson current and the
stability of the $\pi$-state are robust against the ordinary reflection due to the exchange field up to very large spin polarizations $\eta^*$. We obtain a sharp crossover between (i) a regime where the current is given by the quasiclassical theory (ii) the fully spin polarized regime with zero current.    

\section{Acknowledgements}
We thank Igor Zutic for useful comments.

\section{Appendix}\label{appendix1}
In a first approximation, the spectrum of the clean SFS junction is obtained by
a shift $a=2 E_{ex} d/\hbar v_F$ of the phase and it is the solution of the eigenvalue equation

\begin{equation}\label{app1}
\cos \left( \sig a - 2 \varphi_{\ep_o^\sig}  \right) =  \cos \chi
\end{equation}
The exact position of an Andreev level may be written as $\ep^{\sig}(\chi)=\ep_o^\sig (\chi) + \ep_1^\sig (\chi)$, where $\ep_1^\sig (\chi)$ is small. Inserting this expression in Eqs. (\ref{specsfs},\ref{phaseeff}) and using Eq. (\ref{app1}), we obtain for small $\eta$

\begin{eqnarray}\label{app2}
&&\cos \left( \sig a - 2 \varphi_{\ep_o^\sig } + \eta \frac{2 \ep_1^\sig}{E_{\chi}}  \right)
 =  \cos \chi  \\ && -\frac{\eta^2}{8} \cos \chi
 + \frac{\eta^2}{4}
\left(\frac{\ep_o^\sig}{\Delta} \right)^2 \cos 2 k_F d  - \frac{\eta^2}{8} \cos a  \nonumber
\end{eqnarray}
where we have introduced the notation
\begin{equation}
1/E_\chi =\frac{1}{\sqrt{\Delta^2 - {\ep_0^\sig}^2}}.
\end{equation}
Expanding Eq. (\ref{app2}) and using Eq. (\ref{app1}), one obtains a second
order equation for the deviation $\ep_1 (\chi)$

\begin{eqnarray}\label{app3}
&& \cos \chi \left( \frac{\ep_1^\sig}{E_\chi} \right)^2 + \sin \chi \frac{\ep_1^\sig}{E_\chi}
 =  \nonumber \\
 \frac{\eta^2}{16} && \left[ \cos \chi
 -  2 \left( \frac{\ep_o^\sig}{\Delta} \right)^2 \cos 2 k_F d  + \cos a  \right] \nonumber
\end{eqnarray}

For $\chi=0$ and $\chi=\pi$, the deviation is of order $\eta$, whereas for $\chi=\pi/2$ it 
is proportional to $\eta^2$. The gaps occur at the level crossings of the unperturbated spectrum $\ep_o(\chi)$, at $\chi=0$ and $\chi=\pi$. They
are defined by
\begin{eqnarray}\label{app3b}
\delta_0 &=&  \valeurabsolue{\ep_1^\sig (\chi=0) - \ep_1^{-\sig} (\chi=0) }, \nonumber \\
\delta_\pi &=& \valeurabsolue{\ep_1^\sig (\chi=\pi) - \ep_1^{-\sig} (\chi=\pi) }  .\nonumber
\end{eqnarray}
with
\begin{eqnarray}\label{app4}
\frac{\ep_1^\sig (\chi=0)}{E_{0}}&=& \frac{\eta}{2}  \left[ 1 - 2 \left(\frac{\ep_o^\sig}{\Delta} \right)^2 \cos 2 k_F d  + \cos a \right]^{1/2} , \nonumber \\
\frac{\ep_1^\sig (\chi=\pi)}{E_{\pi}} &=& \frac{\eta}{2} \left[ 1 + 2 \left(\frac{\ep_o^\sig}{\Delta} \right)^2 \cos 2 k_F d  - \cos a  \right]^{1/2}  .\nonumber
\end{eqnarray}

Using
\begin{eqnarray}\label{app6}
E_{0} &=&  \Delta   \valeurabsolue{\sin \frac{a}{2}} ,                 \nonumber \\
E_{\pi}  &=&  \Delta \valeurabsolue{\cos \frac{a}{2}}, \nonumber
\end{eqnarray}
we obtain the size of the gaps at $\chi=0$ and $\chi=\pi$
\begin{eqnarray}\label{app5}
\delta_0 &=& \frac{\eta \Delta}{2}  \valeurabsolue{\,\sin k_F d \, \sin a} ,\nonumber \\
\delta_\pi  &=&\frac{\eta \Delta}{2} \valeurabsolue{\,\cos k_F d \, \sin a}. \nonumber \\
\end{eqnarray}


\begin{thebibliography}{99}

\bibitem{ff64} P. Fulde and R.A. Ferrell, Phys. Rev. {\bf 135}, 550 (1964).

\bibitem{lo65} A. I. Larkin and Y.N. Ovchinnikov, Sov. Phys. JETP {\bf 20}, 762 (1965).

\bibitem{bulaev77} L.N. Bulaevskii, V. V. Kuzii and A.A. Sobyanin, Sov. Phys. JETP Lett. {\bf 25}, 290 (1977).

\bibitem{ryazanov01} V.V. Ryazanov, V.A. Oboznov, A.Yu. Rusanov, A.V. Veretennikov, A.A. Golubov and J. Aarts, Phys. Rev. Lett. {\bf 86}, 2427 (2001).

\bibitem{kontos01} T. Kontos, M. Aprili, J. Lesueur and X. Grison, Phys. Rev. Lett. {\bf 86}, 304 (2001).

\bibitem{andreev64} A. F. Andreev, Sov. Phys. JETP {\bf 19}, 1228
(1964).

\bibitem{beenakker92} C.W.J. Beenakker, Phys. Rev. B {\bf 46}, 12841 (1992).

\bibitem{jong95} M.J.M de Jong and C.W.J. Beenakker, Phys. Rev. Lett. {\bf 74}, 1657 (1995).

\bibitem{zutic99} I. Zutic and S. Das Sarma, Phys. Rev. B. {\bf 60}, R16322 (1999).

\bibitem{zutic00} I. Zutic and O.T. Valls, Phys. Rev. B. {\bf 61}, 1555 (2000).

\bibitem{upadhyay98} S.K. Upadhyay, A. Palanisami, R.N. Louie and R.A. Buhrman, Phys. Rev. Lett. {\bf 81}, 3247 (1998).

\bibitem{soulen98} R.J. Soulen, Jr.J.M. Byers, M.S. Osofsky, B. Nadgorny, T. Ambrose, S.F. Cheng, P.R. Broussard, C. T. Tanaka, J. Nowak, J.S. Moodera, A. Barry and  J.M.D. Coey, Science {\bf 282}, 85 (1998).

\bibitem{halterman01} K. Halterman and O.T. Valls, Phys. Rev. B. {\bf 65}, 014509 (2001); cond-mat/0404058.

\bibitem{eschrig03} M. Eschrig, J. Kopu, J.C. Cuevas, and G. Sch\"on, Phys. Rev. Lett. {\bf 90}, 137003 (2003).

\bibitem{buzdin82} A.I. Buzdin, L.N. Bulaevskii and S.V. Panyukov,
JETP Lett. {\bf 35}, 178 (1982).

\bibitem{eilen68} G. Eilenberger, Z. Phys. {\bf 214}, 195 (1968).

\bibitem{chtchel01} N.M. Chtchelkatchev, W. Belzig, Yu.V. Nazarov and C. Bruder,
JETP Lett. {\bf 74}, 323 (2001).

\bibitem{kuple90} S.V. Kuplevakhskii and I.I. Fal'ko,
JETP Lett. {\bf 52}, 343 (1990).

\bibitem{footquasi} "quasiclassical" refers usually to the 3d-quasiclassical theory of superconductivity 
also called the energy integrated Green function's theory in which the fast spatial oscillations
of the Gorkov Green's functions have been integrated out. The resulting quasiclassical Green's functions depend 
only on the center of mass coordinates. In this framework, the current is expressed as a sum of contributions corresponding to different straight line trajectories labelled by a angle $\tta$. In this article, we discuss the purely one-dimensional case and "quasiclassical" refers to the $\tta=0$ contribution of the full quasiclassical result. 

\bibitem{meservey73} P.M. Tedrow and R. Meservey, Phys. Rev. B. {\bf 7}, 318 (1973).

\bibitem{kulik69} I.O. Kulik, Sov. Phys. JETP {\bf 30}, 944
(1970).

\bibitem{bardeen69} J. Bardeen and R. Kümmel and A.E. Jacobs and L. Tewordt, Phys. Rev. {\bf 187}, 556 (1969).

\bibitem{Bardeen72} J. Bardeen and J.L. Johnson, Phys. Rev. B
{\bf5}, 72 (1972).

\bibitem{KOpropre1} I.O. Kulik and A.N. Omel'yanchuk, Sov. J. Low Temp. Phys. {\bf 3}, 459
(1977)

\bibitem{Beenakker1} C.W.J. Beenakker
and H. van Houten, Phys. Rev. Lett. {\bf 66}, 3056 (1991).

\bibitem{Bagwell} P.F. Bagwell, Phys. Rev. B
{\bf46}, 12573 (1992).

\end{thebibliography}
\end{document}